\documentclass[conference]{IEEEtran}
\IEEEoverridecommandlockouts
\usepackage{cite}
\usepackage{amsmath,amssymb,amsfonts}
\usepackage{algorithmic}
\usepackage{graphicx}
\usepackage{textcomp}
\usepackage{xcolor}
\usepackage[caption=false,font=footnotesize,labelfont=rm,textfont=sf]{subfig}
\def\BibTeX{{\rm B\kern-.05em{\sc i\kern-.025em b}\kern-.08em
    T\kern-.1667em\lower.7ex\hbox{E}\kern-.125emX}}
\begin{document}

\title{In-Memory Massive MIMO Linear Detector Circuit with Extremely High Energy Efficiency and Strong Memristive Conductance Deviation Robustness
%{\footnotesize \textsuperscript{*}Note: Sub-titles are not captured in Xplore and should not be used}
\thanks{This work was supported by the Beijing Municipal Natural Science Foundation under Grant L242013. \textit{(Corresponding author: Shaoshi Yang)}}
\thanks{J.-H. Bi, S. Yang and P. Zhang are with the School of Information and Communication Engineering, Beijing University of Posts and Telecommunications, Beijing 100876, China (e-mails: bijiahui@bupt.edu.cn, shaoshi.yang@bupt.edu.cn, pzhang@bupt.edu.cn).}
\thanks{S. Chen is with the School of Electronics and Computer Science, University of Southampton, Southampton SO17 1BJ, U.K. (e-mail: sqc@ecs.soton.ac.uk).}
}

\author{\IEEEauthorblockN{Jia-Hui Bi, Shaoshi~Yang,~\IEEEmembership{Senior Member,~IEEE}, Ping~Zhang,~\IEEEmembership{Fellow,~IEEE}, Sheng Chen,~\IEEEmembership{Life Fellow,~IEEE}}}

\maketitle

\begin{abstract}
The memristive crossbar array (MCA) has been successfully applied to accelerate matrix computations of signal detection in massive multiple-input multiple-output (MIMO) systems. However, the unique property of massive MIMO channel matrix makes the detection performance of existing MCA-based detectors sensitive to conductance deviations of memristive devices, and the conductance deviations are difficult to be avoided. In this paper, we propose an MCA-based detector circuit, which is robust to conductance deviations, to compute massive MIMO zero forcing and minimum mean-square error algorithms. The proposed detector circuit comprises an MCA-based matrix computing module, utilized for processing the small-scale fading coefficient matrix, and amplifier circuits based on operational amplifiers (OAs), utilized for processing the large-scale fading coefficient matrix. We investigate the impacts of the open-loop gain of OAs, conductance mapping scheme, and conductance deviation level on detection performance and demonstrate the performance superiority of the proposed detector circuit over the conventional MCA-based detector circuit. The energy efficiency of the proposed detector circuit surpasses that of a traditional digital processor by several tens to several hundreds of times.
\end{abstract}

\begin{IEEEkeywords}
Massive MIMO, signal detection, linear detector, analog matrix computing, in-memory computing, memristive crossbar array.
\end{IEEEkeywords}

\section{Introduction}\label{S1}

Modern and next-generation wireless communication systems employ massive multiple-input multiple-output (MIMO) technology to increase transmission speed and improve user experience. However, the extremely large number of antennas, while beneficial, also leads to extremely-high complexity of signal detection algorithms. With the goal of reducing detection latency in massive MIMO systems, a variety of low-complexity detection algorithms have been proposed in the past decades \cite{Survey_50years}. However, algorithms with low complexity usually suffer from considerable performance loss, making it difficult to trade off between high performance and low latency. Another popular and effective approach, as exemplified in previous studies \cite{traditional1,traditional2}, is to accelerate MIMO detection by hardware innovations. However, traditional processors based on the von Neumann architecture struggle significantly to perform large-dimensional matrix operations. As the number of users simultaneously transmitting data increases in next-generation wireless communication systems, the computational complexity of detection algorithms is bound to increase, and traditional von Neumann architecture-based processors are difficult to meet the requirements of receiver for processing speed and energy efficiency.

As a form of in-memory computing, the analog matrix computing technology based on memristive crossbar array (MCA) constitutes a revolutionary new matrix computational paradigm. The MCA can rapidly perform not only matrix-vector multiplication (MVM) \cite{MVMref} through analog computing approach but also other matrix operations, such as the computation of inverse matrix \cite{SunZhong_Inv} and pseudoinverse matrix \cite{SunZhong_GeneralizedInv}, with the assistance of operational amplifiers (OAs). The MCA-based matrix computing circuit is not constrained by the so-called von Neumann bottleneck, and thereby presents notable benefits in computational speed and power consumption in contrast to traditional processors based on the von Neumann architecture. The MCA makes it possible to realize massive MIMO detectors with superior detection performance, speed, and energy efficiency, since that matrix operations constitute the core task of massive MIMO detection.

While MCA has been successfully applied to accelerate deep neural network training in artificial intelligence, research on the application of it in massive MIMO detection is still in its infancy. {\color{black}In \cite{Memristor_baseband_processors}, MCA was applied to baseband processors to accelerate the minimum mean-square error (MMSE) based MIMO detection. But this work only employed MCA to accelerate the MVM operations, and relied on another processor to perform inverse matrix computations. A zero forcing (ZF) precoder circuit, which was constructed by an MCA-based MVM module and an MCA-based inversion module, was proposed in \cite{Memristor_Precoder}. This idea can be extended to the ZF detector. Two MCA-based detector circuits with similar structures were proposed in \cite{Memristor_MIMO_Acceleration} and \cite{Realizing_InMemory_Baseband}, respectively, and both circuits can perform linear detection algorithms, including ZF, regularized ZF and MMSE algorithms, in one step. In \cite{YiHangRen}, MCA was employed to accelerate the maximum likelihood detector.} However, in practical scenarios, the large-scale fading coefficients (LSFCs) of the user terminals (UTs) in a massive MIMO cell usually vary from each other, and thus the elements of the matrices computed in MCA-based circuits presented in \cite{Memristor_baseband_processors,Memristor_Precoder,Memristor_MIMO_Acceleration, Realizing_InMemory_Baseband} often obey probability distributions with different variances, which makes the detection performance of the existing MCA-based detectors sensitive to conductance deviations.

To solve this problem, we propose an MCA-based detector circuit in this paper, which can be employed to compute massive MIMO ZF and MMSE algorithms and is robust to conductance deviations. The proposed detector circuit comprises an MCA-based matrix computing module, utilized for processing the small-scale fading coefficient (SSFC) matrix, and OA-based amplifier circuits, utilized for processing the LSFC matrix. Compared with conventional MCA-based detector circuit, the proposed circuit exhibits smaller differences concerning the values of the elements in the mapped matrix. As a result, the proposed circuit demonstrates strong memristive conductance deviation robustness, leading to better detection performance than conventional MCA-based detector circuit. We also demonstrate the energy efficiency superiority of the proposed circuit over the traditional digital processor.

\section{System Model and Basic Algorithms}\label{S2}

\subsection{System Model}\label{S2.1}

We consider a massive MIMO cell, where a base station (BS) with $R$ antennas serves $K$ UTs, each equipped with a single antenna. The uplink signals can be described by:
\begin{equation}\label{y=Hs+n_Complex} % eq.1
	\tilde{\bf{y}} = \tilde{\bf{H}} \tilde{\bf{s}} + \tilde{\bf{n}}, 
\end{equation}
where $\tilde{\bf{y}} \in \mathbb{C}^{R\times 1}$ denotes the received signals, $\tilde{\bf{s}}\! \in\! \mathbb{C}^{K\times 1}$ denotes the transmitted signals, $\tilde{\bf{H}}\! \in\! \mathbb{C}^{R\times K}$ denotes the channel matrix, and $\tilde{\bf{n}}\! \in\! \mathbb{C}{^{R\times 1}}$ is a complex additive white Gaussian noise (AWGN) vector with variance $\sigma_n^2$ per element.

Let $\lambda_1, \cdots ,\lambda_K$ be the LSFCs between the $K$ UTs and the BS, then $\tilde{\bf{H}}$ can be described by:
\begin{equation}\label{eqChMa} % eq.2
	\tilde{\bf{H}} = \tilde{\bf{G}} \tilde{\bf{\Lambda}},
\end{equation}
where $\tilde{\bf{\Lambda}}=\textrm{diag}\big(\sqrt{\lambda _1}, \cdots ,\sqrt{\lambda_K}\big)$ and ${\tilde{\bf{G}}} \in \mathbb{C}^{R\times K}$ are the LSFC matrix and the SSFC matrix, respectively. We consider the Rayleigh fading channel model in this paper, which means that the elements of $\tilde{\bf{G}}$ are zero-mean complex Gaussian random variables with variance $\sigma_g^2$ per dimension, i.e., 
\begin{equation}\label{eqRaFa} % eq.3
  \tilde{g}_{i,j} \sim {\cal CN}\big(0,2\sigma_g^2\big) , \, 1\le i\le R, \, 1\le j\le K .
\end{equation}

The complex-valued system model of \eqref{y=Hs+n_Complex} can be alternatively described as an equivalent real-valued expression of
\begin{equation}\label{eqRVmimo} % eq.4
	\bf{y} = \bf{H} \bf{s} + \bf{n},
\end{equation}
where 
\begin{align*}
  & \bf{y} = \left[ \begin{array}{c}
  \Re \big( \tilde{\bf{y}} \big) \\
  \Im \big( \tilde{\bf{y}} \big)
  \end{array} \right] , ~
  \bf{s} = \left[ \begin{array}{c}
  \Re \big( \tilde{\bf{s}} \big) \\
  \Im ( \tilde{\bf{s}} \big)
  \end{array} \right] , ~
  \bf{n} = \left[ \begin{array}{c}
  \Re \big( \tilde{\bf{n}} \big) \\
  \Im \big( \tilde{\bf{n}} \big)
  \end{array} \right] , \\
  & \bf{H} = \left[ \begin{array}{cc}
  \Re \big( \tilde{\bf{H}} \big) & -\Im \big( \tilde{\bf{H}} \big) \\
  \Im \big( \tilde{\bf{H}} \big) &  \Re \big( \tilde{\bf{H}} \big)
  \end{array} \right],
\end{align*}  
in which $\Re (\cdot )$ and $\Im (\cdot )$ respectively denote the real and imaginary parts of the corresponding vector or matrix. Obviously, ${\bf{H}}\in \mathbb{R}^{2R\times 2K}$ can be alternatively described by:
\begin{equation}\label{H=GLamda} % eq.5
	\bf{H} = \bf{G} \bf{\Lambda},
\end{equation}
where ${\bf{\Lambda}}=\textrm{diag}\big(\sqrt{\lambda_1}, \cdots ,\sqrt{\lambda_{K}}, \sqrt{\lambda_1}, \cdots ,\sqrt{\lambda_{K}}\big)$ and
\begin{align*}
  \bf{G} =& \left[ \begin{array}{cc}
  \Re \big( \tilde{\bf{G}} \big) & -\Im \big( \tilde{\bf{G}} \big) \\
  \Im \big( \tilde{\bf{G}} \big) &  \Re \big( \tilde{\bf{G}} \big)
\end{array} \right] .
\end{align*}

A massive MIMO detector needs to estimate $\bf{s}$ from $\bf{y}$ given $\bf{H}$.

\subsection{Basic Detection Algorithms}\label{S2.2}
\subsubsection{ZF Algorithm}

The ZF algorithm can be expressed as:
\begin{equation}\label{ZF_Real} % eq.6
	\hat{\bf{s}}_{\textrm{ZF}} = \big({\bf{H}}^{\textrm T} {\bf{H}}\big)^{-1} {\bf{H}}^{\textrm T} \bf{y} ,
\end{equation}
where $(\cdot )^{\textrm T}$ and $(\cdot )^{-1}$ denote the transpose matrix and inverse matrix, respectively.

\subsubsection{MMSE Algorithm}

The MMSE algorithm can be expressed as:
\begin{equation} \label{MMSE_Real} % eq.8
	\hat{\bf{s}}_{\textrm{MMSE}} = \big({\bf{H}}^{\textrm T} {\bf{H}} + \rho {\bf{I}} \big)^{-1} {\bf{H}}^{\textrm T} {\bf{y}} ,
\end{equation}
where the parameter $\rho=\frac{\sigma_{n}^{2}}{p_s}$, $p_s$ is the average symbol energy of $\bf{s}$, and $\bf{I}$ donotes the identity matrix of appropriate dimension.

\section{Proposed MCA-Based Circuit Design}\label{S3}

\subsection{Transformations of Computational Expressions}

Upon substituting \eqref{H=GLamda} into \eqref{ZF_Real} and \eqref{MMSE_Real} we obtain:
\begin{equation}\label{eqZF} % eq.7
  \hat{\bf{s}}_{\textrm{ZF}} = {\bf{\Lambda}}^{-1}\big({\bf{G}}^{\textrm T} {\bf{G}}\big)^{-1} {\bf{G}}^{\textrm T} {\bf{y}},
\end{equation}
and
\begin{equation}\label{eqMMSE} % eq.9
	\hat{\bf{s}}_{\textrm{MMSE}} = {\bf{\Lambda}}^{-1}\big({\bf{G}}^{\textrm T} {\bf{G}} + {\bf{P}}\big)^{-1} {\bf{G}}^{\textrm T} \bf{y} ,
\end{equation}
where ${\bf{P}}=\textrm{diag}\big(\frac{\rho}{\lambda_1}, \frac{\rho}{\lambda_2}, \cdots, \frac{\rho}{\lambda_{K}}, \frac{\rho}{\lambda_1}, \frac{\rho}{\lambda_2}, \cdots, \frac{\rho}{\lambda_{K}}\big)$.

For expression convenience, we define ${\bf{W}}={\bf{G}}^{\textrm T}{\bf{G}}$, and $\tilde{\bf{W}}=\tilde{\bf{G}}^{\textrm H}\tilde{\bf{G}}$, where $(\cdot )^{\textrm H}$ denotes the Hermitian transpose. Thus $\Re \big( {\tilde{\bf{W}}} \big)$ and $\Im \big( {\tilde{\bf{W}}} \big)$ are two real symmetric matrices and $\bf{W}$ can be expressed as:
\begin{equation}\label{eqProINV} % eq.16
	{\bf{W}} = \left[ {\begin{array}{*{20}{c}}
    {\Re \big( {\tilde{\bf{W}}} \big)} & {-\Im \big( {\tilde{\bf{W}}} \big)}\\
    {\Im \big( {\tilde{\bf{W}}} \big)} & {\Re \big( {\tilde{\bf{W}}} \big)}
  \end{array}} \right].
\end{equation}

The real and imaginary parts of the elements ${\tilde{g}}_{i,j}$ of $\tilde{\bf{G}}$ are independent identically distributed Gaussian random variables, namely, $\Re\big({\tilde{g}}_{i,j}\big)\! \sim\! {\cal N}(0,\sigma _g^2)$ and $\Im\big({\tilde{g}}_{i,j}\big)\! \sim\! {\cal N}(0,\sigma _g^2)$. Thus the mean values of the nondiagonal elements of both $\Re \big( {\tilde{\bf{W}}} \big)$ and $\Im \big( {\tilde{\bf{W}}} \big)$ are zeros, and the diagonal elements of $\Im \big( {\tilde{\bf{W}}} \big)$ are zeros, while the diagonal elements of $\Re \big( {\tilde{\bf{W}}} \big)$ obey a chi-square distribution:
\begin{equation}\label{eqChi} % eq.17
  \frac{\Re({\tilde{w}}_{i,i})}{\sigma _g^2} \sim \chi ^2(2R),
\end{equation}
which means that the mean value of the diagonal elements of $\Re \big( {\tilde{\bf{W}}} \big)$ is $2R\sigma_g^2$.

By defining ${\bf{Q}}_{\textrm{ZF}}\! =\! 2R\sigma_g^2 {\bf{I}}$, ${\bf{Q}}_{\textrm{MMSE}}\! =\! 2R\sigma_g^2 {\bf{I}}\! +\! {\bf{P}}$ and ${\bf{X}}\! =\! {\bf{W}} - 2R\sigma_g^2 {\bf{I}}$, we obtain:
\begin{equation}\label{eqProINV1} % eq.18
	{\hat{\bf{s}}}_{\textrm{ZF}} = {\bf{\Lambda}}^{-1}\big({\bf{X}} + {\bf{Q}}_{\textrm{ZF}}\big)^{-1} {\bf{G}}^{\textrm T} {\bf{y}},
\end{equation}
and
\begin{equation}\label{eqProINV2} % eq.19
	{\hat{\bf{s}}}_{\textrm{MMSE}} = {\bf{\Lambda}}^{-1}\big({\bf{X}} + {\bf{Q}}_{\textrm{MMSE}}\big)^{-1} {\bf{G}}^{\textrm T} {\bf{y}}.
\end{equation}

\subsection{Proposed MCA-Based Circuit}

\begin{figure}[tp]
  \vspace*{-1mm}
    \centerline{\includegraphics[width=\linewidth]{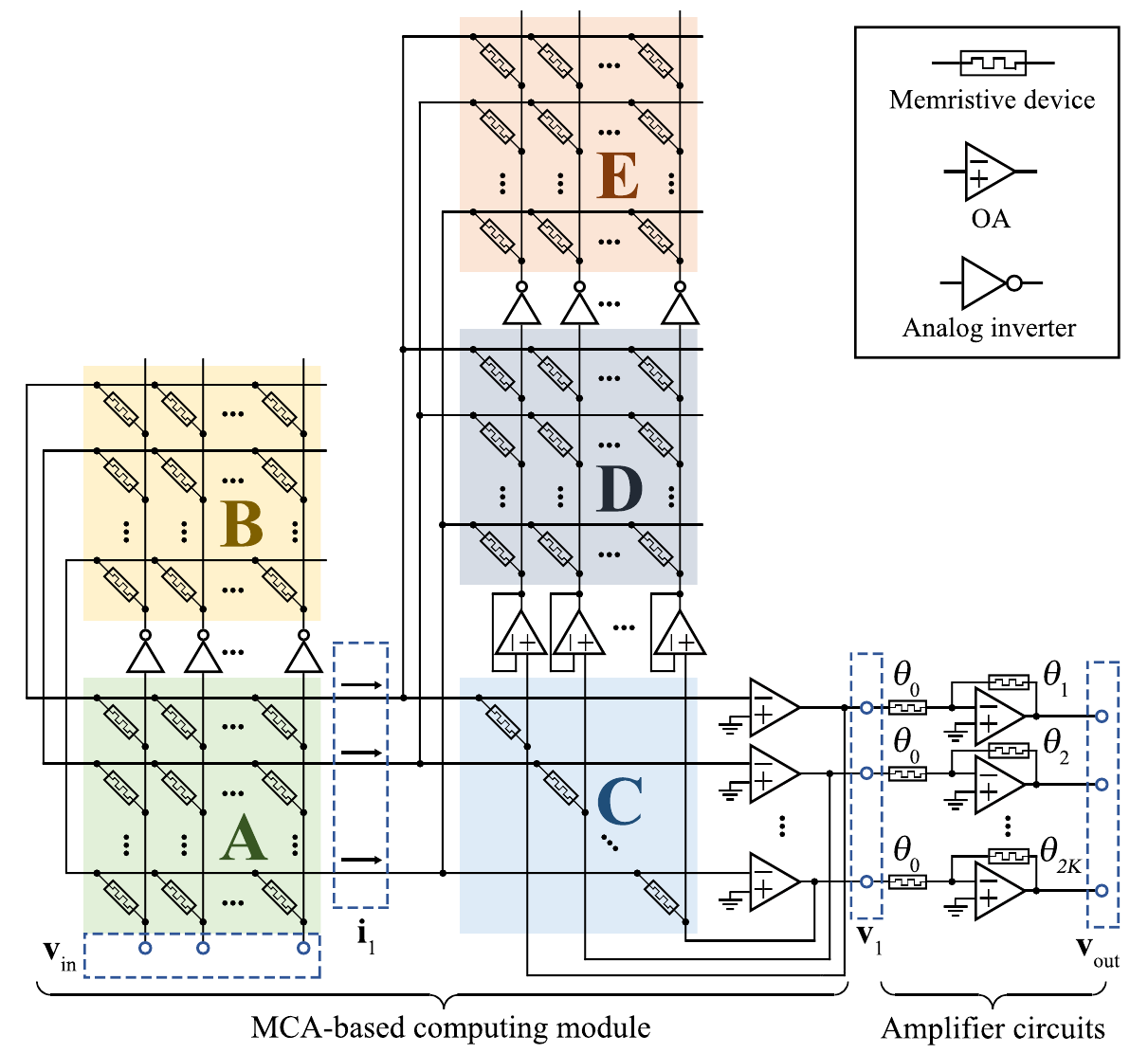}}
  \vspace*{-2mm}
   \caption{The proposed MCA-based detector circuit.}
  \label{detector} % Fig.3
  \vspace*{-4mm}
\end{figure}

The proposed detector circuit is illustrated in Fig.~\ref{detector}, which comprises an MCA-based computing module and $2K$ amplifier circuits. The MCA-based computing module consists of five MCAs, two sets of analog inverters, a set of voltage followers and a set of OAs.

Owing to the virtual ground property of OA networks, the voltages at the inverting-input nodes of the set of OAs are approximately zeros. Let $\bf{A}$, $\bf{B}$, $\bf{C}$, $\bf{D}$ and $\bf{E}$ be the conductance matrices of the five MCAs. According to Ohm's law and Kirchhoff's law, the input voltages ${\bf{v}}_{\textrm{in}}$ and the currents ${\bf{i}}_{1}$ in Fig.~\ref{detector} satisfy:

\begin{equation}\label{INV_1} % eq.20
  {\bf{i}}_{1} = ({\bf{A}} - {\bf{B}}) {\bf{v}}_{\textrm{in}}.
\end{equation}

A voltage follower has a unity-gain. Therefore, let ${\bf{v}}_{1}$ be the output voltages of the set of OAs, we have:

\begin{equation}\label{INV_2} % eq.11
	({\bf{C}} + {\bf{D}} - {\bf{E}}){\bf{v}}_{1} + {\bf{i}}_{1}=  {\bf{i}}^{-} ,
\end{equation}
where ${\bf{i}}^{-}$ denotes the currents flowing into the inverting-input nodes of OAs. Since ${\bf{i}}^{-}$ is approximately zeros owing to the inherent characteristic of OAs, we have
\begin{equation}\label{INV_3} % eq.12
	{\bf{v}}_{1} = -({\bf{C}} + {\bf{D}} - {\bf{E}})^{-1} {\bf{i}}_{1}.
\end{equation}
The stability of the output voltages requires that the signs of the diagonal elements of ${\bf{C}}^{-1}$ are all positive \cite{SunZhong_Inv}, which is always valid, since ${\bf{C}}$ is a diagonal matrix with positive diagonal elements.

In the amplifier circuits, the conductance values of the memristive devices connected to the output nodes of the MCA-based computing module are all $\theta _0$. Let $\theta_1, \theta_2, \cdots, \theta_{2K}$ be the conductance values of the feedback memristive devices, respectively. The output voltages of the amplifier circuits are:
\begin{equation}\label{INV_4} % eq.22
  {\bf{v}}_{\textrm{out}} = -{\bf{\Theta}}^{-1} {\bf{v}}_{1},
\end{equation}
where ${\bf{\Theta}}=\textrm{diag}\big(\frac{\theta_1}{\theta _0}, \frac{\theta_2}{\theta_0}, \cdots, \frac{\theta_{2K}}{\theta_0}\big)$. Upon substituting \eqref{INV_1} and \eqref{INV_3} into \eqref{INV_4}, we obtain:

\begin{equation}\label{INV_5} % eq.23
  {\bf{v}}{_{\textrm{out}}} = {\bf{\Theta}}^{-1} ({\bf{C}} + {\bf{D}} - {\bf{E}})^{-1} ({\bf{A}}-{\bf{B}}) {\bf{v}}_{\textrm{in}}.
\end{equation}

The conductance value of a memristive device can be changed by charge or flux through it. Therefore, the conductance value of a memristive device can be set to any desired value within a specified range by a dedicated program \cite{Analogue_VMM}. By mapping ${\bf{y}}$ onto ${\bf{v}}_{\textrm{in}}$, mapping ${\bf{G}}^{\textrm T}$ onto ${\bf{A}}-{\bf{B}}$, mapping ${\bf{Q}}_{\textrm{ZF}}$ or ${\bf{Q}}_{\textrm{MMSE}}$ onto ${\bf{C}}$, mapping ${\bf{X}}$ onto ${\bf{D}}-{\bf{E}}$ and mapping ${\bf{\Lambda}}$ onto ${\bf{\Theta}}$, the result of \eqref{eqProINV1} or \eqref{eqProINV2}, i.e., $\hat{\bf{s}}_{\textrm{ZF}}$ or $\hat{\bf{s}}_{\textrm{MMSE}}$, can be obtained by measuring ${\bf{v}}_{\textrm{out}}$. 

\section{Conductance Mapping Schemes}\label{S4} % S4

The accuracy of mapping the matrices to be computed to the conductance matrices directly impacts the accuracy of matrix computation. This mapping accuracy is determined by both the mapping scheme and the conductance deviations of memristive devices (i.e., the difference between the target and actual conductances of the devices). In this section, we give two conductance mapping schemes, namely, the fixed mapping factor (FMF) scheme and the adjustable mapping factor (AMF) scheme.

We map a matrix that contains both negative and positive elements onto the difference between two positive conductance matrices, rather than a single one, in order to align with physical constraint. Let the conductance range of memristive devices be $[\omega_{\textrm{min}}, ~ \omega_{\textrm{max}}]$. Let $\bf{U}$ be the mapped matrix, i.e., ${\bf{G}}^{\textrm T}$ or ${\bf{X}}$, and let ${\bf{A}}$ and ${\bf{B}}$ be the corresponding conductance matrices. The scheme for mapping $\bf{U}$ onto ${\bf{A}}-{\bf{B}}$ is:
\begin{equation}\label{eqS4-1} % eq.29
  a_{i,j} = \begin{cases}
    {\omega_{\textrm{max}},u_{i,j} > 0} \\
    {\omega_{\textrm{min}},u_{i,j} \leq  0}
    \end{cases}
\end{equation}
and
\begin{equation}\label{eqS4-2} % eq.30
  b_{i,j} = a_{i,j} - \alpha u_{i,j},
\end{equation}
where $\alpha$ is called the mapping factor. The conductance values exceeding the permissible range will be truncated to the limits.

\subsection{FMF Scheme}\label{S4.1}

The core concept of the FMF scheme is to select a fixed mapping factor based on the probability distribution of elements of the mapped matrix. To map a matrix $\bf{U}$ onto conductance matrices, the FMF scheme calculates the mapping factor by:
\begin{equation}\label{eqS4-3} % eq.31
  \alpha = \frac{\omega}{\beta\sigma_u},
\end{equation}
where $\omega = \omega_{\textrm{max}} - \omega_{\textrm{min}}$, $\beta$ is a parameter of the FMF scheme and $\sigma_u$ is the standard deviation of the elements of $\bf{U}$.

\subsection{AMF Scheme}\label{S4.2}

The AMF scheme calculates the mapping factor by:
\begin{equation}\label{eqS4-7} % eq.35
  \alpha=\frac{\omega}{{\textrm{max}}\{\left|u_{i,j}\right|\}},
\end{equation}
to map $\bf{U}$ onto conductance matrices.

\section{Simulations}\label{S5} % S5

To evaluate the proposed MCA-based detector circuit, we simulate a $4 \times 64$ massive MIMO system, employing 64-quadrature-amplitude-modulation (64-QAM). The conventional MCA-based detection scheme computes $\hat{\bf{s}}_{\textrm{ZF}}$ or $\hat{\bf{s}}_{\textrm{MMSE}}$ according to \eqref{ZF_Real} or \eqref{MMSE_Real}, so we use the MCA-based computing module in Fig.~\ref{detector} to represent the conventional MCA-based detector circuit in our experiments. We consider memristive devices whose conductance range is $0.1 \rm{\mu}$S $ \sim 30 \rm{\mu}$S, and we perform simulations using the circuit simulator LTspice$^\circledR$.

\begin{figure}[bp]
\vspace*{-6mm}
  \centering
  \subfloat[]{\includegraphics[width=0.5\linewidth]{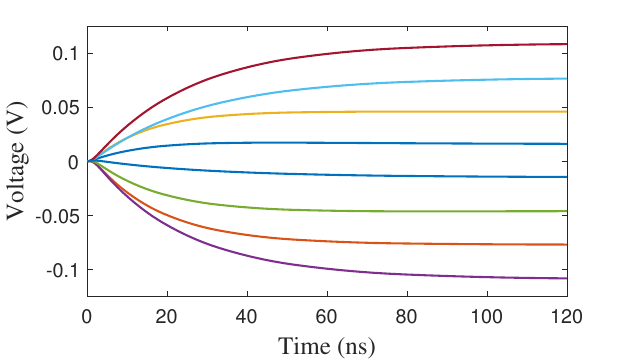}}
  \hfil
  \subfloat[]{\includegraphics[width=0.5\linewidth]{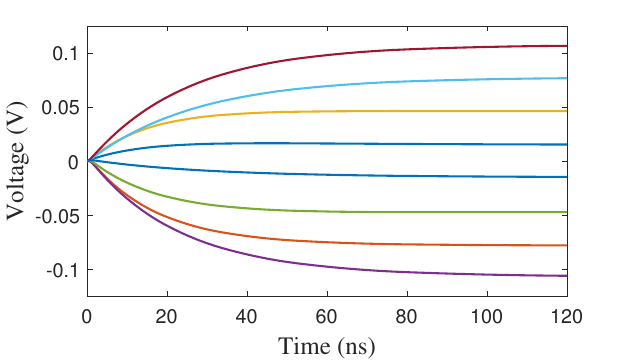}}
\vspace*{-1mm}
\caption{Waveforms of output voltages: (a)~the proposed detector circuit, and (b)~the conventional MCA-based detector circuit.}
\label{time} % Fig.7
\end{figure}

\subsection{Computation Time}

We gauge the computation time of an MCA-based detector circuit by its convergence time, which is mainly influenced by the gain-bandwidth product (GBP) of OAs \cite{Convergence_Time}. The output voltage waveforms of the proposed detector circuit and the conventional MCA-based detector circuit are shown in Fig.~\ref{time}, and the OAs are assumed to have a GBP of 500\,MHz. The convergence time of the proposed circuit is about 110\,ns, exhibiting almost no difference compared with that of the conventional MCA-based detector circuit, and can be further enhanced by increasing the GBP of OAs.

\subsection{Detection Performance}

In this subsection, we first investigate the impacts of the OLG of OAs, conductance mapping scheme and conductance deviations on detection performance of the proposed circuit, and then demonstrate the performance superiority of the proposed circuit over the conventional MCA-based detector circuit. The conductance deviations are modeled as zero-mean Gaussian random variables with a variance of $\sigma_m^2$ \cite{Error}.

\begin{figure}[tbp]
  \centerline{\includegraphics[width=\linewidth]{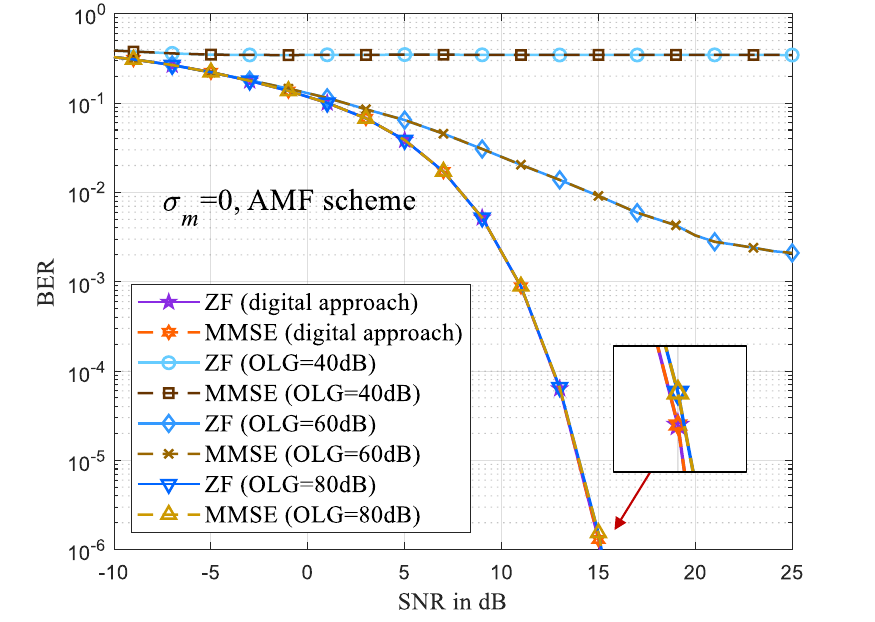}}
  \vspace*{-3mm}
  \caption{BERs of the proposed detector circuit under different values of the OLG of OAs when $\sigma_m=0$ and adopting the AMF scheme.}
  \label{BER_alpha0}
  \vspace*{-3mm}
\end{figure}

The computational accuracy of an MCA-based detector circuit is significantly constrained by the OLG of OAs. Fig.~\ref{BER_alpha0} shows the bit error rate (BER) results as the functions of the signal-to-noise ratio (SNR) for the proposed detector circuit, given various values of the OLG of OAs with $\sigma_m=0$ and adopting the AMF scheme, in comparison with the BER of the digital benchmark. When the OLG of OAs is too low, the detection performance is poor. The OLG of OAs needs to be at least 80\,dB for the proposed detector circuit to ensure satisfactory performance, i.e., achieving the performance of the digital benchmark. In the rest of this subsection, we assume that the OLG of OAs is sufficiently large.

\begin{figure}[tbp]
  \centerline{\includegraphics[width=\linewidth]{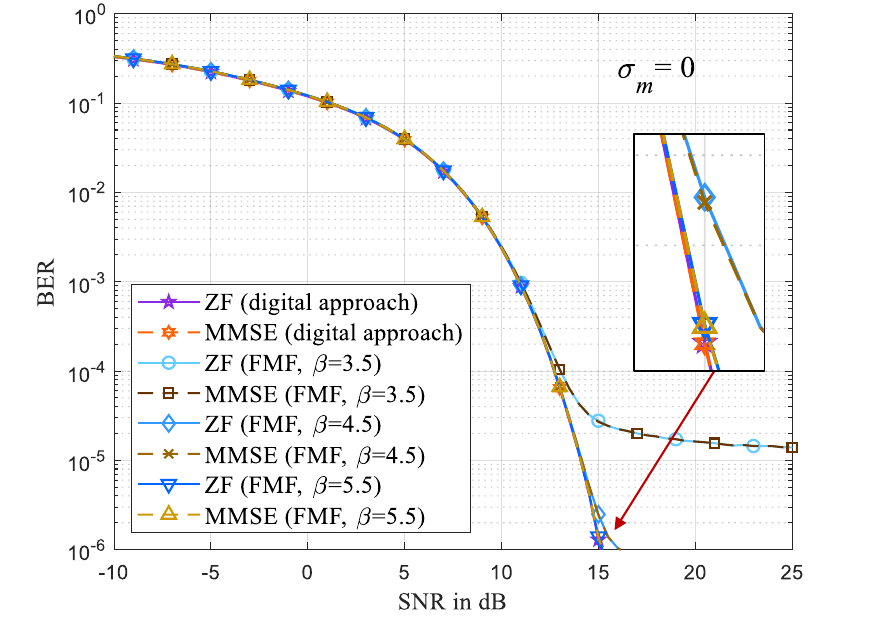}}
  \vspace*{-3mm}
  \caption{BERs of the proposed detector circuit under different values of $\beta$ when $\sigma_m=0$ and adopting the FMF scheme.}
  \label{BERvsSNR}
  \vspace*{-3mm}
\end{figure}

Fig.~\ref{BERvsSNR} shows the BER results of the proposed detector circuit adopting the FMF scheme under different $\beta$ values with $\sigma_m=0$, again using the digital approach as the benchmark. Obviously, when $\sigma_m=0$, a higher value of $\beta$ results in fewer elements being truncated, and thus results in a lower BER and closer performance to the digital approach for the proposed detector circuit. Although the truncated elements degrade the detection performance, such an impact is mainly noticeable in high SNR. In low SNR, however, the primary constraint on detection performance remains the AWGN. Even without AWGN, detection errors still occur due to the truncated elements, causing the BER to gradually converge to a fixed value as the SNR increases.

\begin{figure}[tbp]
  \centerline{\includegraphics[width=\linewidth]{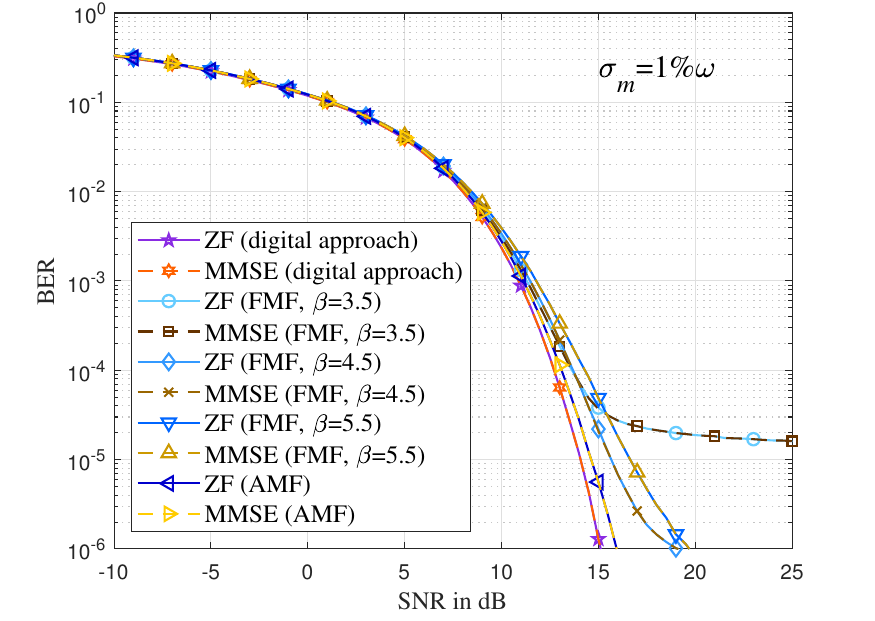}}
  \vspace*{-3mm}
  \caption{BERs of the proposed detector circuit with $\sigma_m=1$\%$\omega$.}
  \label{BERvsSNR_2}
  \vspace*{-3mm}
\end{figure}

Fig.~\ref{BERvsSNR_2} shows the BER results of the proposed detector circuit, with $\sigma_m=1\% \omega$. Obviously, the BER decreases and then increases as $\beta$ increases, because the primary factor constraining detection performance shifts from the truncated elements to the perturbations caused by conductance deviations as $\beta$ increases. {\color{black}Observe from Fig. \ref{BER_alpha0} $\sim$ Fig. \ref{BERvsSNR_2} that in the considered scenarios there is no noteworthy difference in the detection performance between ZF and MMSE algorithms. Hence in what follows the performance of the MMSE algorithm will be used to represent the linear detectors' performance.}

\begin{figure}[bp]
  \vspace*{-5mm}
  \centerline{\includegraphics[width=0.95\linewidth]{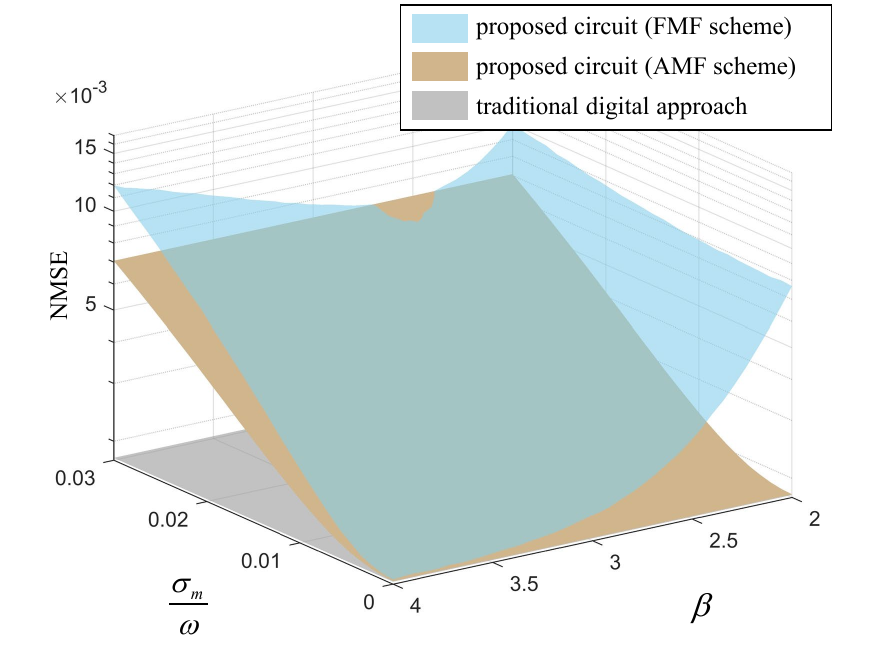}}
  \vspace*{-3.5mm}
  \caption{NMSEs of the computational results relative to the transmitted signals of the proposed MCA-based detector circuit, adopting the FMF scheme and the AMF scheme, given an SNR of 20 dB and using the digital approach as the benchmark.}
  \label{SRE3D}
\end{figure}

To facilitate the observation of the performance differences among the detector circuits with similar BERs, we use the normalized mean squared error (NMSE) of the computational results relative to the transmitted signals to measure the detection performance of a detector circuit. Fig.~\ref{SRE3D} depicts the NMSEs of the computational results of the proposed detector circuit and the conventional detector circuit as the functions of both conductance deviation level and $\beta$, given an SNR of 20 dB and using the digital approach as the benchmark. For the FMF scheme, in the absence of conductance deviation, the larger the parameter $\beta$, the smaller the NMSE, while in the presence of conductance deviations, the NMSE first decreases and then increases as $\beta$ increases, which confirms the trend observed in Fig.~\ref{BERvsSNR_2}. Besides, only when the conductance deviation level is high and an appropriate $\beta$ is selected, the performance of the FMF scheme surpasses that of the AMF scheme, otherwise it is worse than that of the AMF scheme.

\begin{figure}[tbp]
  \centerline{\includegraphics[width=\linewidth]{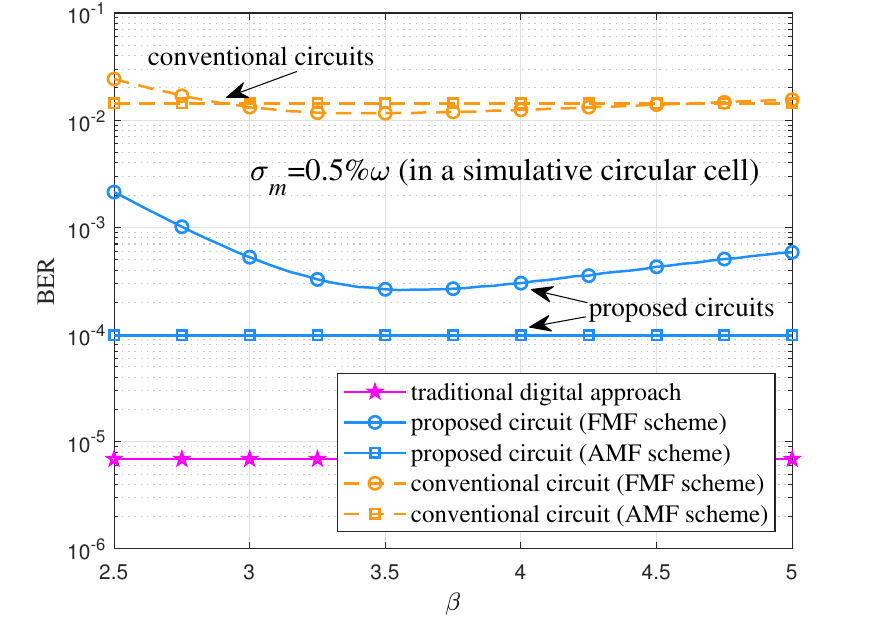}}
  \vspace*{-3mm}
  \caption{BERs of the proposed detector circuit and the conventional detector circuit, varying with $\beta$ value in a massive MIMO cell, with $\sigma_m=0.5\%\omega$.}
  \label{contrast}
  \vspace*{-3mm}
\end{figure}

In practical scenarios, UTs in a cell are usually located at different positions, the disparity in LSFCs associated with UTs results in great differences in the variances of the different elements of the matrices computed in the conventional MCA-based detector circuit, and the perturbations caused by conductance deviations are particularly severe to the elements with smaller variance. For the proposed detector circuit, the elements of $\bf{G}$ obey the same distribution, although the diagonal and nondiagonal elements of $\bf{X}$ follow different distributions, their variance disparity is not significant. Consequently, the impact of conductance deviations on the performance of the proposed detector circuit is relatively minor compared to that on the conventional MCA-based detector circuit. 

To showcase the performance superiority of our proposed circuit over the conventional MCA-based detector circuit, we consider a multi-user massive MIMO cell whose radius is 150\,m. UTs randomly appear within the cell, with each UT has a transmitting power of 20\,dBm. The carrier frequency of uplink signals is 2\,GHz with a bandwidth of 25\,MHz. Fig.~\ref{contrast} shows the BER results of the proposed detector circuit and the conventional MCA-based detector circuit varying with $\beta$ value, with $\sigma_m=0.5\%\omega$. Obviously, regardless of whether the AMF or FMF mapping scheme is employed, the proposed detector circuit consistently achieves a notably lower BER compared to the conventional MCA-based detector circuit.

\subsection{Power Consumption and Energy Efficiency}

\begin{figure}[tbp]
  \centerline{\includegraphics[width=\linewidth]{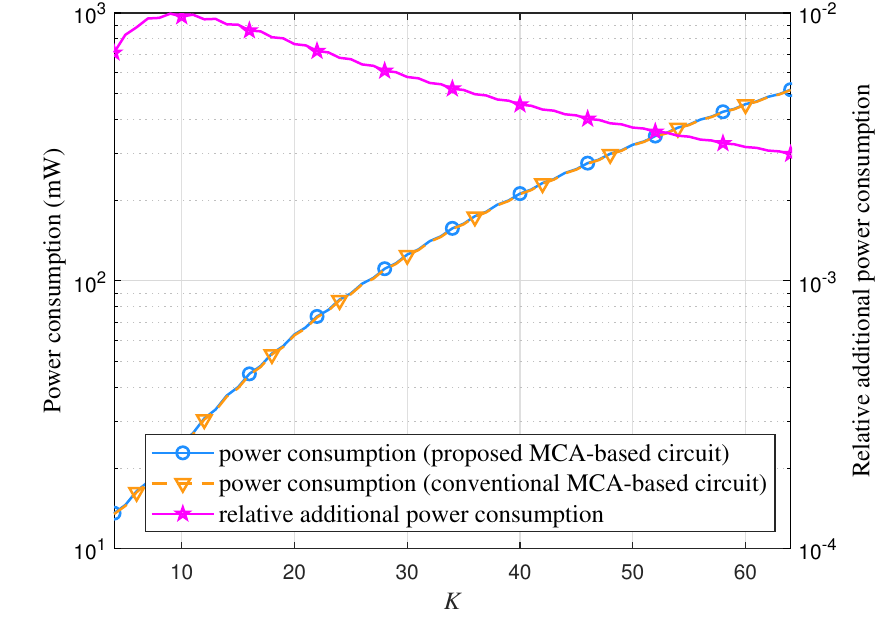}}
  \vspace*{-3mm}
  \caption{Power consumption results of the proposed circuit and the conventional MCA-based detector circuit, as well as the relative additional power consumption of the proposed circuit, varying with the number of UTs, $K$.}
   \label{power} % Fig.16
  \vspace*{-3mm}
\end{figure}

Fig.~\ref{power} shows the estimated power consumption results of the proposed circuit and the conventional MCA-based detector circuit, while varying $K$. In this experiment, we consider the OA of \cite{OAref}. Digital-to-analog converters (DACs) of \cite{VDACref} and analog-to-digital converters (ADCs) of \cite{ADC_Array} are employed in our experiments to supply input voltages to the circuits and measure output voltages of the circuits, respectively. Fig.~\ref{power} also shows the relative additional power consumption of the proposed circuit, i.e., the ratio of the power consumption of the introduced amplifier circuits to that of the conventional MCA-based detector circuit, which is less than 1\%.

\begin{figure}[tbp]
\centerline{\includegraphics[width=\linewidth]{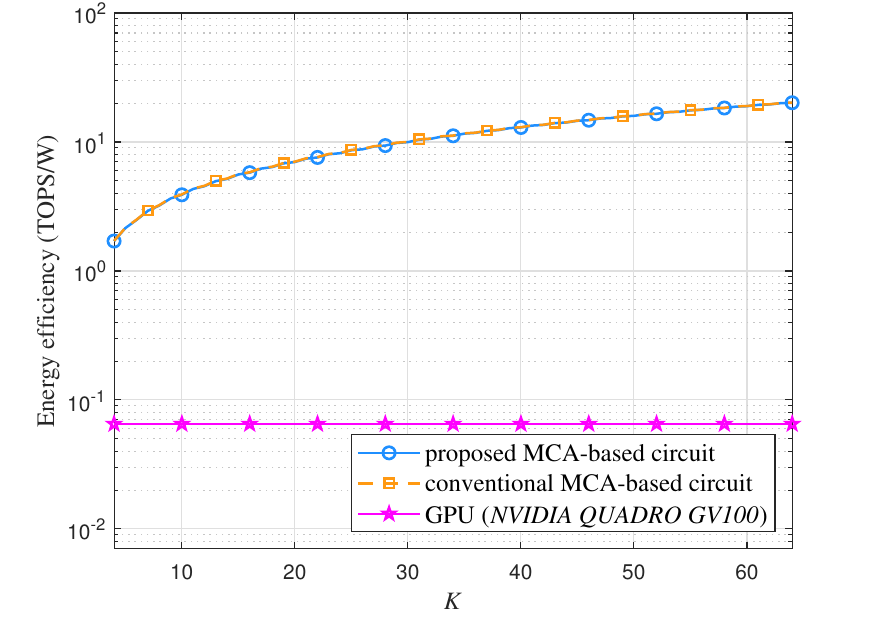}}
\vspace*{-3mm}
\caption{Energy efficiency results of the proposed detector circuit, the conventional MCA-based detector circuit and a GPU benchmark.}
\label{EE} % Fig.18
\vspace*{-3mm}
\end{figure}

The energy efficiency of an MCA-based detector circuit can be gauged by the ratio of its equivalent floating-point operation (FLOP) number to the energy consumed during its computation time, which is measured in tera-FLOPs per second per watt (TOPS/W) in this paper. It is noted that either a real multiplication or a real summation is considered as a FLOP. Fig.~\ref{EE} shows the energy efficiency results of the proposed circuit, the conventional MCA-based detector circuit and the commercial graphic processing unit (GPU) {\it{NVIDIA QUADRO GV100}} \cite{GPU}. The energy efficiency of the proposed circuit is almost identical to that of the conventional MCA-based detector circuit. As the number of UTs increases, the energy efficiency of the MCA-based detector circuits also increases, which is several orders of magnitude higher than that of the GPU {\it{NVIDIA QUADRO GV100}}.

\section{Conclusions}\label{Conclusion}

In this paper, we have proposed an MCA-based detector circuit, which can be employed to compute massive MIMO ZF and MMSE algorithms. In contrast to all existing MCA-based detector circuits, our proposed detector circuit comprises an MCA-based matrix computing module, utilized for processing the SSFC matrix, and OA-based amplifier circuits, utilized for processing the LSFC matrix, and thereby achieves high robustness against conductance deviations of the memristive devices. We have investigated the impacts of the OLG of OAs, conductance mapping scheme, and conductance deviation level on detection performance of the proposed detector circuit. The proposed detector circuit exhibits significant performance superiority compared to the conventional MCA-based detector circuit, only incurring a negligible additional cost of power consumption. Moreover, the energy efficiency of our proposed circuit is several orders of magnitude higher than that of the commercial GPU {\it{NVIDIA QUADRO GV100}}.

\end{document}